\documentclass[%
onecolumn,
superscriptaddress,
showpacs,
preprintnumbers,
showkeys,
nofootinbib,
 amsmath,amssymb,
 aps,
 prd,
floatfix
]{revtex4-1}
\usepackage{amsthm}
\usepackage{mathtools}
\usepackage{physics}
\usepackage{xcolor}
\usepackage{graphicx}
\usepackage[paperwidth=210mm,paperheight=297mm,centering,hmargin=2cm,vmargin=2.5cm]{geometry}
\usepackage{adjustbox}
\usepackage{placeins}
\usepackage[T1]{fontenc}
\usepackage{lipsum}
\usepackage{csquotes}

\usepackage{xspace}

\newcommand{\minerva}{MINERvA\xspace}

\usepackage{listings} % inline code
\usepackage{xcolor} % inline code color

\definecolor{codegreen}{rgb}{0,0.6,0}
\definecolor{codegray}{rgb}{0.5,0.5,0.5}
\definecolor{codepurple}{rgb}{0.58,0,0.82}
\definecolor{codeblue}{rgb}{0,0.3,0.6}
\definecolor{backcolour}{rgb}{0.95,0.95,0.92}
\definecolor{dkgreen}{rgb}{0,0.6,0}
\definecolor{dred}{rgb}{0.545,0,0}
\definecolor{dblue}{rgb}{0,0,0.545}
\definecolor{lgrey}{rgb}{0.9,0.9,0.9}
\definecolor{gray}{rgb}{0.4,0.4,0.4}
\definecolor{darkblue}{rgb}{0.0,0.0,0.6}

\lstdefinestyle{mystyle}{
  language=C++,
  backgroundcolor=\color{backcolour},   commentstyle=\color{codegreen},
  numberstyle=\tiny\color{codegray},
  stringstyle=\color{codepurple},
  basicstyle=\ttfamily\footnotesize,
  breakatwhitespace=false,         
  breaklines=true,                 
  captionpos=b,                    
  keepspaces=true,                 
  numbers=left,                    
  numbersep=5pt,                  
  showspaces=false,                
  showstringspaces=false,
  showtabs=false,                  
  tabsize=2,
  keywordstyle = {\color{dblue}},
  keywordstyle = [2]{\color{dkgreen}},
  keywordstyle = [3]{\color{codepurple}},
  keywordstyle = [4]{\color{codeblue}},
  otherkeywords = {;,<<,>>,++},
  morekeywords = [2]{PlotUtils},
  morekeywords = [3]{GetFluxSystematics},
  morekeywords = [4]{container_of_systematics},
  stringstyle = {\color{dred}},
  breakatwhitespace,
}

\newcommand{\universe}{\lstinline[style=myStyle]|Universe|\xspace}
\newcommand{\universes}{\lstinline[style=myStyle]|Universes|\xspace}
\newcommand{\mnvhnd}{\lstinline[style=myStyle]|MnvHnD|\xspace}
\newcommand{\mnvhnds}{\lstinline[style=myStyle]|MnvHnDs|\xspace}
\newcommand{\mnvhoned}{\lstinline[style=myStyle]|MnvH1D|\xspace}
\newcommand{\thnd}{\lstinline[style=myStyle]|THnD|\xspace}
\newcommand{\thoned}{\lstinline[style=myStyle]|TH1D|\xspace}
\newcommand{\mnverrorband}{\lstinline[style=myStyle]|MnvErrorBand|\xspace}
\newcommand{\mnverrorbands}{\lstinline[style=myStyle]|MnvErrorBands|\xspace}
\newcommand{\passescuts}{\lstinline[style=myStyle]|PassesCuts()|\xspace}
\newcommand{\energyHistogram}{\lstinline[style=myStyle]|energyHistogram|\xspace}
\begin{document}
\title{An Error Analysis Toolkit for Binned Counting Experiments}

%% List of institution addresses, in command form.
%% List of institution addresses, in command form.
\newcommand{\Rutgers}{Rutgers, The State University of New Jersey, Piscataway, New Jersey 08854, USA}
\newcommand{\Hampton}{Hampton University, Dept. of Physics, Hampton, VA 23668, USA}
\newcommand{\Dortmund}{Institute of Physics, Dortmund University, 44221, Germany }
\newcommand{\Otterbein}{Department of Physics, Otterbein University, 1 South Grove Street, Westerville, OH, 43081 USA}
\newcommand{\JMU}{James Madison University, Harrisonburg, Virginia 22807, USA}
\newcommand{\Florida}{University of Florida, Department of Physics, Gainesville, FL 32611}
\newcommand{\UCIrvine}{Department of Physics and Astronomy, University of California, Irvine, Irvine, California 92697-4575, USA}
\newcommand{\CBPF}{Centro Brasileiro de Pesquisas F\'{i}sicas, Rua Dr. Xavier Sigaud 150, Urca, Rio de Janeiro, Rio de Janeiro, 22290-180, Brazil}
\newcommand{\PUCP}{Secci\'{o}n F\'{i}sica, Departamento de Ciencias, Pontificia Universidad Cat\'{o}lica del Per\'{u}, Apartado 1761, Lima, Per\'{u}}
\newcommand{\INRM}{Institute for Nuclear Research of the Russian Academy of Sciences, 117312 Moscow, Russia}
\newcommand{\Jlab}{Jefferson Lab, 12000 Jefferson Avenue, Newport News, VA 23606, USA}
\newcommand{\Pittsburgh}{Department of Physics and Astronomy, University of Pittsburgh, Pittsburgh, Pennsylvania 15260, USA}
\newcommand{\Guanajuato}{Campus Le\'{o}n y Campus Guanajuato, Universidad de Guanajuato, Lascurain de Retana No. 5, Colonia Centro, Guanajuato 36000, Guanajuato M\'{e}xico.}
\newcommand{\Athens}{Department of Physics, University of Athens, GR-15771 Athens, Greece}
\newcommand{\Tufts}{Physics Department, Tufts University, Medford, Massachusetts 02155, USA}
\newcommand{\WM}{Department of Physics, College of William \& Mary, Williamsburg, Virginia 23187, USA}
\newcommand{\FNAL}{Fermi National Accelerator Laboratory, Batavia, Illinois 60510, USA}
\newcommand{\Purdue}{Department of Chemistry and Physics, Purdue University Calumet, Hammond, Indiana 46323, USA}
\newcommand{\MCLA}{Massachusetts College of Liberal Arts, 375 Church Street, North Adams, MA 01247}
\newcommand{\UMD}{Department of Physics, University of Minnesota -- Duluth, Duluth, Minnesota 55812, USA}
\newcommand{\Northwestern}{Northwestern University, Evanston, Illinois 60208}
\newcommand{\UNI}{Facultad de Ciencias, Universidad Nacional de Ingenier\'{i}a, Apartado 31139, Lima, Per\'{u}}
\newcommand{\Rochester}{University of Rochester, Rochester, New York 14627 USA}
\newcommand{\Austin}{Department of Physics, University of Texas, 1 University Station, Austin, Texas 78712, USA}
\newcommand{\USM}{Departamento de F\'{i}sica, Universidad T\'{e}cnica Federico Santa Mar\'{i}a, Avenida Espa\~{n}a 1680 Casilla 110-V, Valpara\'{i}so, Chile}
\newcommand{\Geneva}{University of Geneva, 1211 Geneva 4, Switzerland}
\newcommand{\Chicago}{Enrico Fermi Institute, University of Chicago, Chicago, IL 60637 USA}
\newcommand{\hired}{}
\newcommand{\OregonState}{Department of Physics, Oregon State University, Corvallis, Oregon 97331, USA}
\newcommand{\oxford}{Oxford University, Department of Physics, Oxford, OX1 3PJ United Kingdom}
\newcommand{\umiss}{University of Mississippi, Oxford, Mississippi 38677, USA}
\newcommand{\upenn}{Department of Physics and Astronomy, University of Pennsylvania, Philadelphia, PA 19104}
\newcommand{\AMU}{AMU Campus, Aligarh, Uttar Pradesh 202001, India}
\newcommand{\wroclaw}{University of Wroclaw, plac Uniwersytecki 1, 50-137 Wrocław, Poland}
\newcommand{\Mohali}{Department of Physical Sciences, IISER Mohali, Knowledge City, SAS Nagar, Mohali - 140306, Punjab, India}
\newcommand{\CINVESTAV}{Departamento de Fisica Col. San Pedro Zacatenco, 07360 Mexico, DF, Av. Instituto Politécnico Nacional, Mexico}
\newcommand{\york}{York University, Department of Physics and Astronomy, Toronto, Ontario, M3J 1P3 Canada}
\newcommand{\ND}{Department of Physics, University of Notre Dame, Notre Dame, Indiana 46556, USA}
\newcommand{\ICL}{The Blackett Laboratory,  Imperial College London,  London SW7 2BW, United Kingdom}

%% New affiliations
\newcommand{\benmesserlyThanks}{Now at University of Minnesota}
\newcommand{\robfineThanks}{Now at Los Alamos National Laboratory}
\newcommand{\mateusfcarneiroThanks}{Now at Brookhaven National Laboratory}

% 45 total signatories.
%% Primary authors
\author{B.~Messerly}\thanks{\benmesserlyThanks} \affiliation{\Pittsburgh}
\author{R.~Fine}\thanks{\robfineThanks}         \affiliation{\Rochester}
\author{A.~Olivier}                            \affiliation{\Rochester}
%% Collaboration authors
\author{Z.~~Ahmad~Dar}                    \affiliation{\WM}  \affiliation{\AMU}
\author{F.~Akbar}                         \affiliation{\AMU}
\author{M.~V.~Ascencio}                   \affiliation{\PUCP}
\author{A.~Bashyal}                       \affiliation{\OregonState}
\author{L.~Bellantoni}                    \affiliation{\FNAL}
\author{A.~Bercellie}                     \affiliation{\Rochester}
\author{J.~L.~Bonilla}                    \affiliation{\Guanajuato}
\author{G.~Caceres}                       \affiliation{\CBPF}
\author{T.~Cai}                           \affiliation{\Rochester}
\author{M.F.~Carneiro}\thanks{\mateusfcarneiroThanks}  \affiliation{\OregonState}
\author{G.A.~D\'{i}az~}                   \affiliation{\Rochester}
\author{J.~Felix}                         \affiliation{\Guanajuato}
\author{L.~Fields}                        \affiliation{\FNAL}
\author{A.~Filkins}                       \affiliation{\WM}
\author{A.~Ghosh}                         \affiliation{\USM}  \affiliation{\CBPF}
\author{S.~Gilligan}                      \affiliation{\OregonState}
\author{R.~Gran}                          \affiliation{\UMD}
\author{H.~Haider}                        \affiliation{\AMU}
\author{D.A.~Harris}                      \affiliation{\york}  \affiliation{\FNAL}
\author{S.~Henry}                         \affiliation{\Rochester}
\author{S.~Jena}                          \affiliation{\Mohali}
\author{D.~Jena}                          \affiliation{\FNAL}
\author{J.~Kleykamp}                      \affiliation{\Rochester}
\author{M.~Kordosky}                      \affiliation{\WM}
\author{D.~Last}                          \affiliation{\upenn}
\author{A.~Lozano}                        \affiliation{\CBPF}
\author{X.-G.~Lu}                         \affiliation{\oxford}
\author{K.S.~McFarland}                   \affiliation{\Rochester}
\author{C.~Nguyen}                        \affiliation{\Florida}
\author{V.~Paolone}                       \affiliation{\Pittsburgh}
\author{G.N.~Perdue}                      \affiliation{\FNAL}  \affiliation{\Rochester}
\author{M.A.~Ram\'{i}rez}                 \affiliation{\upenn}  \affiliation{\Guanajuato}
\author{H.~Ray}                           \affiliation{\Florida}
\author{D.~Ruterbories}                   \affiliation{\Rochester}
\author{H.~Schellman}                     \affiliation{\OregonState}
\author{C.J.~Solano~Salinas}              \affiliation{\UNI}
\author{H.~Su}                            \affiliation{\Pittsburgh}
\author{E.~Valencia}                      \affiliation{\WM}  \affiliation{\Guanajuato}
\author{N.H.~Vaughan}                     \affiliation{\OregonState}
\author{B.~Yaeggy}                        \affiliation{\USM}
\author{K.~Yang}                          \affiliation{\oxford}
\author{L.~Zazueta}                       \affiliation{\WM}

\collaboration{The \minerva Collaboration} \email{Please direct correspondence to mess@umn.edu, finer@fnal.gov, and aolivier@ur.rochester.edu} \noaffiliation

\date{\today} % Leave empty to omit a date

\begin{abstract}
    \vspace{5ex}
    We introduce the \minerva Analysis Toolkit (MAT), a utility for centralizing the handling of systematic uncertainties in HEP analyses. The fundamental utilities of the toolkit are the \mnvhnd, a powerful histogram container class, and the systematic \universe classes, which provide a modular implementation of the many universe error analysis approach. These products can be used stand-alone or as part of a complete error analysis prescription. They support the propagation of systematic uncertainty through all stages of analysis, and provide flexibility for an arbitrary level of user customization. This extensible solution to error analysis enables the standardization of systematic uncertainty definitions across an experiment and a transparent user interface to lower the barrier to entry for new analyzers.
\end{abstract}

\maketitle

\section{Introduction}
\label{sec:intro}
Despite the importance and complexity of systematic uncertainty evaluation in HEP, there is no community-wide standard for the storage and propagation of uncertainty through analysis infrastructures. At the experiment level, an error analysis system must be a central consideration during the design of analysis infrastructure. In the absence of such a system, consideration of systematic uncertainty is incumbent on individual analyzers and can be relegated to the final stages of analysis after substantial design choices have already been made. This impedes the effectual integration of error analysis and represents a substantial barrier to entry for new analyzers. To this point, analyzers within an experiment are often left to implement solutions to this independently of one another, creating an inefficiency of effort and increased opportunity for inconsistency and mistakes.

We present a solution to this problem, the \minerva Analysis Toolkit (MAT), which can be adopted to streamline an experiment's approach to evaluating systematic uncertainty. The MAT (1) implements the many-universe error method, (2) is built on a unique and powerful histogram object (\mnvhnd), and (3) offers experiment-wide standardization of systematics and simultaneous customizability and extensibility via \universe classes. This toolkit was developed within the \minerva analysis environment where it has facilitated dozens of neutrino cross section measurements over the last decade \cite{Filkins:2020xol,Coplowe:2020yea,Carneiro:2019jds,Cai:2019hpx,Le:2019jfy,Valencia:2019mkf,Stowell:2019zsh,Elkins:2019vmy,Ruterbories:2018gub,Lu:2018stk,Gran:2018fxa,Patrick:2018gvi,Mislivec:2017qfz,Altinok:2017xua,Betancourt:2017uso,Ren:2017xov,Marshall:2016yho,DeVan:2016rkm,Aliaga:2016oaz,Wang:2016pww,McGivern:2016bwh,Marshall:2016rrn,Wolcott:2016hws,Mousseau:2016snl,Park:2015eqa,Rodrigues:2015hik,Wolcott:2015hda,Aliaga:2015wva,Walton:2014esl,Higuera:2014azj,Eberly:2014mra,Tice:2014pgu,Fiorentini:2013ezn,Fields:2013zhk} and forms the basis for \minerva's data preservation effort \cite{Fine:2020snd}. The MAT was designed to be adopted directly, in part or in whole, by any HEP experiment. It is directly applicable to any ROOT-based \cite{Brun:1997pa} analysis environment in which simulated physics interactions are evaluated on an event-by-event basis. We also believe that the underlying approach and some of the classes which underpin the MAT's functionality could be implemented in any software environment for a binned counting experiment in any field of study.

 In this paper, we introduce in detail the design of the MAT, provide examples for its current and future use within \minerva, and describe how it could be adopted more widely. Section \ref{sec:systematics} introduces the "many universe" approach to evaluating systematic uncertainty, which is a prerequisite for this toolkit. Sections \ref{sec:mnvhnd} and \ref{sec:universes_technical} describe the technical implementations of the \mnvhnd and \universe classes, and describe how MAT analyses are structured, including pseudocode examples. Finally, Section \ref{sec:application} discusses MAT integration into different experimental analysis environments and benefits of adopting the MAT wholesale.

\vspace{-1ex}
\section{"Many Universe" Error Analysis}
\label{sec:systematics}
%P1
Modern experimental HEP is inextricably coupled to complex simulations that are designed to replicate the physics we measure. These simulations are parameterized by variables that correspond to the design of an experiment and the related physics processes. In general, the output of these simulations is a collection of individual interactions (events), which are subject to the same reconstruction and analysis used for data. The observables of an analysis are binned as functions of kinematic variables and stored in histograms. When analyzed in aggregate over many events, the distribution of simulated event kinematics can be compared to data to judge the efficacy of the simulation. In this manner data is used to fine-tune the simulation and ultimately to improve the underlying physics models.

%P2
In the "many universe" approach to evaluating systematic uncertainty, aspects of the simulation and reconstruction are shifted to reflect distinct sources of uncertainty. An analysis is repeated many times, and in each instance the effect of a particular shift is propagated through subsequent stages of the simulation-reconstruction-analysis chain.  The particular configuration of the simulation and reconstruction in any individual instance defines a "universe". Two universes may be distinct from one another, for example, because of the choice of values for parameters within the beamline simulation or within a track reconstruction algorithm. The choice of nominal values (or "best guesses") for all parameters in the simulation and reconstruction defines the "Central Value" (CV) universe and forms the basis against which data measurements are compared, and against which uncertainties are evaluated. Each systematic universe is constructed to isolate the impact of shifting a particular aspect of the CV universe in a measured way, and the uncertainty on a measurement corresponding to each individual source is determined based on the spread of outcomes across two or more universes.

%P3
It can be constructive to distinguish between two conceptually different classes of shifts away from the CV. The first, \textit{vertical shifts}, arise from sources of systematic uncertainty which do not directly affect any kinematic variable, and their effect is only to modify the weight a particular event is given in an analysis. Vertical shifts cause the contents of a histogram bin to be increased or decreased but do not lead to event migration between bins, and therefore will never lead to events migrating into or out of a selected sample. Because of this, the effect of a vertical shift of a far-upstream aspect of the simulation can be characterized downstream in the analysis by changing the relative weighting of events based on their kinematics. This "reweighting" enables evaluation of vertical systematic shifts without the need to repeat computationally intensive stages of the simulation. The second, \textit{lateral shifts}, arise from sources of systematic uncertainty that do directly affect a kinematic variable and therefore can cause events to migrate between bins and into and out of a selected sample. A lateral shift may cause an event which does not pass a selection cut in the CV universe to instead pass in the shifted universe, or vice versa. The impact of a lateral shift on the final result must be determined by repeating the portion of the simulation-reconstruction-analysis chain downstream of the shift, and cannot be evaluated via reweighting.

%P4
In practice, the value of a systematic shift is determined \textit{a priori}, and often corresponds to the standard deviation of a Gaussian distribution. For example, a dedicated study of muon momenta might inform a $\pm2\%$ shift applied to muons on an event-by-event basis to represent the uncertainty associated with their reconstruction. In this case, the impact of this uncertainty on an observable is calculated using the variance between the $+2\%$- and $-2\%$-shifted universes in each bin of the analysis. Many uncertainties are adequately represented by two universes because the pair of universes reflects a simple underlying distribution (usually assumed to be Gaussian). Others motivate a large number of universes to encapsulate the effect of simultaneously varying multiple parameters within a more complex model. For example, neutrino beamline fluxes are predicted using notoriously complex simulations, in which the correlations among internal parameters are nontrivial. In this case, an ensemble of universes is used, and in each multiple parameters are varied according to their individual statistical distributions. The impact of this uncertainty on an observable is then calculated using the variance between all of the universes in each bin of the analysis.

%P5
Within the MAT, individual systematic universes are represented by \universe classes, and universes corresponding to the same source of uncertainty are grouped into "error bands". The \mnvhnd class, described in detail in the following section, provides the interface to propagate an analysis histogram in many systematic universes simultaneously. The uncertainty on an observable stored in an \mnvhnd can be calculated on-demand at any stage of the analysis by using the spread across universes to construct a covariance matrix corresponding to any individual systematic uncertainty. The covariance matrix corresponding to any individual systematic uncertainty, or to the total uncertainty, can be reported, or the diagonal entries of the covariance matrix can be used to produce vertical error bars on a histogram. This approach to reporting uncertainty is common in HEP, but the many universe technique does not exclude alternate error treatments.

\vspace{-1ex}
\section{The MnvHnD Class}
\label{sec:mnvhnd}
A successful implementation of the many universe approach requires that an analysis be repeated many times. In the MAT, this is achieved using an \mnvhnd object, which stores the central value and systematic universe histograms for an analysis observable. With all histograms stored in a single object, analysis calculations can proceed simultaneously in all universes. The interface of the \mnvhnd is implemented by broadcasting the \thnd interface from ROOT to each systematic universe, so uncertainties are automatically propagated correctly when \mnvhnds are \lstinline[style=myStyle]{Add()}ed, \lstinline[style=myStyle]{Divide()}d, or \lstinline[style=myStyle]{Scale()}d. An \mnvhnd can \lstinline[style=myStyle]{Write()} itself to a ROOT file to persist an analysis result with everything needed to produce a covariance matrix, defer further processing to a separate program, or separate plot formatting from computationally intensive analysis routines.
  
\begin{figure}
  \centering
  \includegraphics[width=0.49\textwidth]{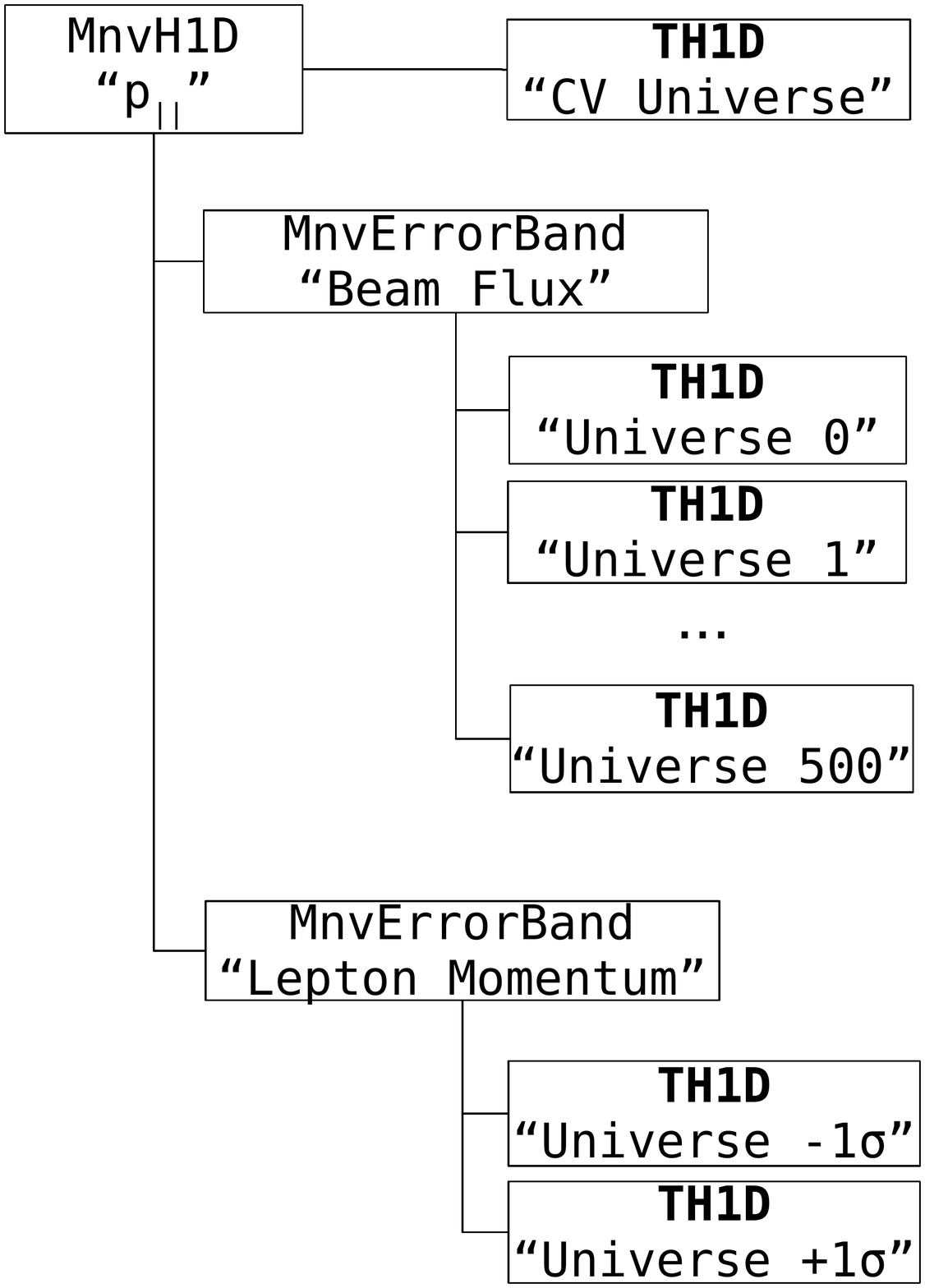}
  \includegraphics[width=0.49\textwidth]{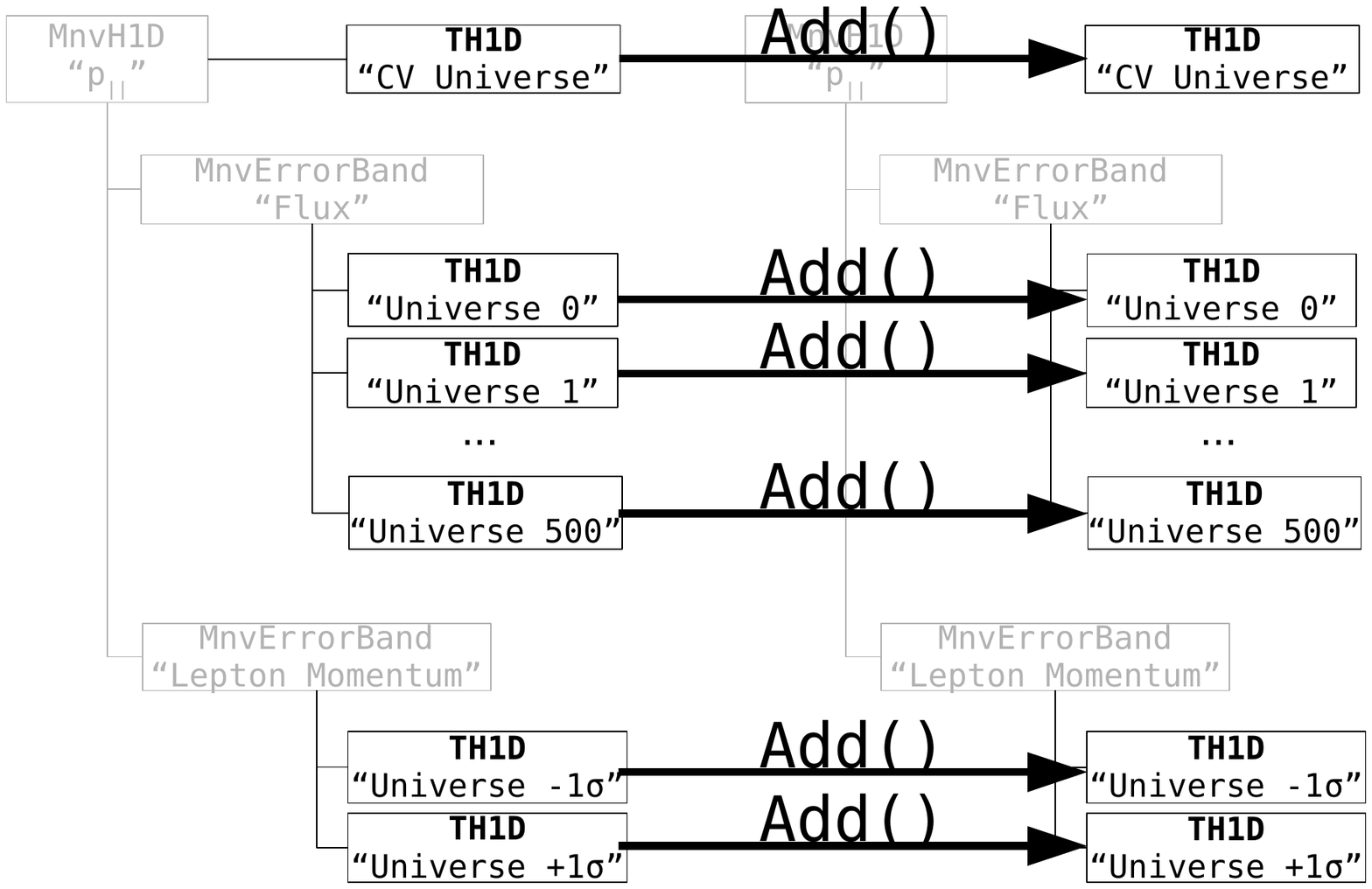}
  \caption[Structure of MnvH1D]{Left: An \mnvhoned encapsulates one histogram for each \universe.  \universes are grouped into \mnverrorbands.  The \mnvhoned is itself the CV \thoned.  Right: The \mnvhoned propagates uncertainties by performing \thoned operations like \lstinline[style=myStyle]{Add()} in each \universe.}
  \label{fig:MnvH1DStructure}
\end{figure}

An \mnvhnd encapsulates a parallel histogram for each systematic universe as shown in Figure \ref{fig:MnvH1DStructure}. To populate a histogram in a particular universe, an analyzer need only tell an \mnvhnd which \universe a \lstinline[style=myStyle]{Fill()} value belongs to (as in line 12 of Code Example \ref{lst:eventLoop}). The covariance matrix for any group of \universes can then be calculated on demand. An uncertainty is usually evaluated using two or more \universes grouped into an error band (implemented in the MAT as the \mnverrorband class). Covariance matrices are calculated for each error band individually, and the total covariance matrix on a result is equal to the sum of all of the individual covariance matrices. In this way, the variance of each bin is added in quadrature as expected for uncorrelated uncertainties.
  
%%%%%%%%%%%%%%%%%%%%%%%
%%%%%%%%%%%%%% Figure 1
%%%%%%%%%%%%%%%%%%%%%%%
%%%%%%%%%%%%%%%%%%%%%%%

\begin{figure}
	\centering
	\includegraphics[width=0.49\textwidth]{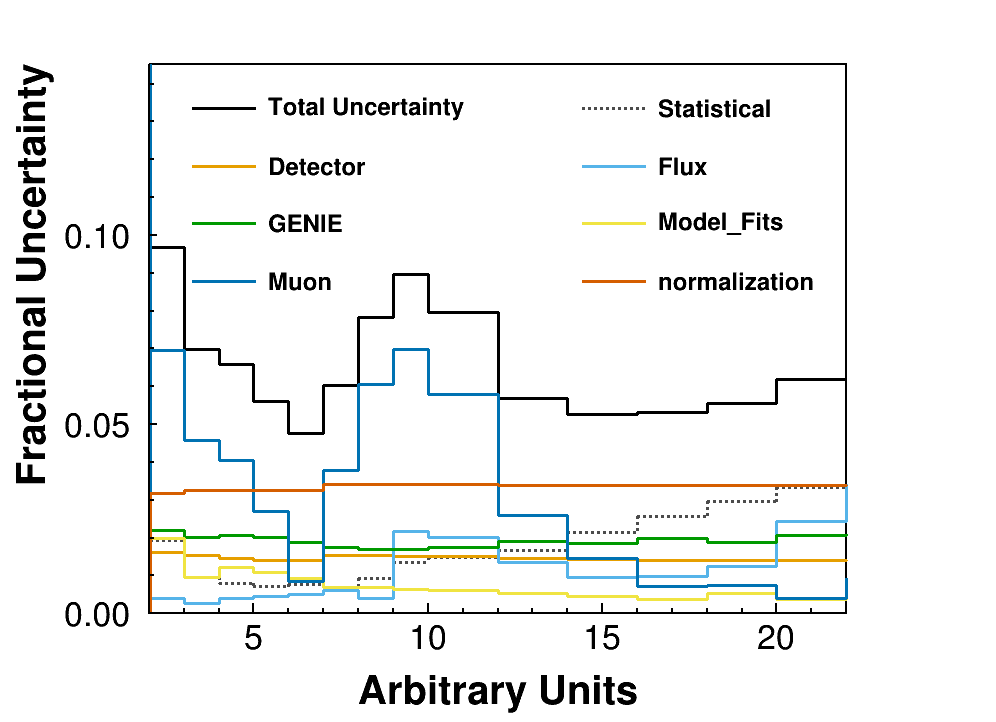}
	\includegraphics[width=0.49\textwidth]{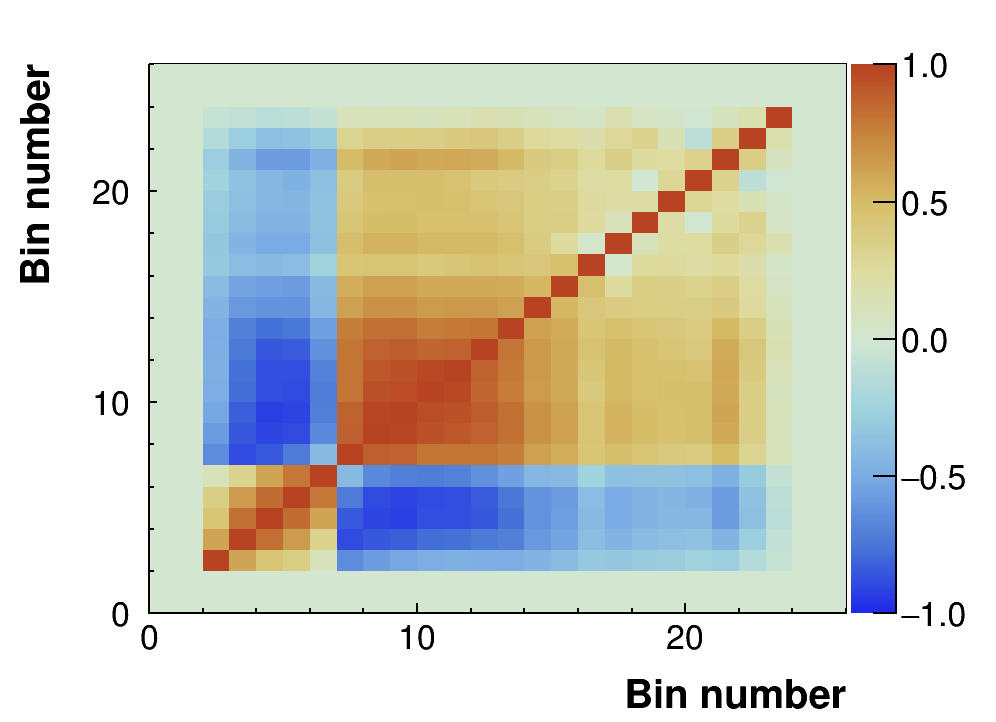}
	\begin{lstlisting}[style=myStyle]
    hist->DrawErrorSummary();      // [left]
    errMatrix = hist->GetCorrelationMatrix(); 
    errMatrix.Draw();              // [right]
\end{lstlisting}
	\caption[Example Correlation Matrix]{The MAT provides data visualization tools that operate on an \mnvhnd. Left: An error budget is constructed by extracting the diagonal entries of distinct \mnverrorbands' covariance matrices to show the relative contributions of independent sources of uncertainty. Right: The total correlation matrix is constructed by adding together (and normalizing) the covariance matrices of all \mnverrorbands.}
	\label{fig:systematicsBreakdown}
\end{figure}

The MAT provides tools to visualize the data and uncertainties stored in an \mnvhnd. Because each error band is propagated independently, it is straightforward to see how each source of uncertainty contributes to the total uncertainty of a distribution. Figure \ref{fig:systematicsBreakdown} shows an example of an uncertainty breakdown as a function of an arbitrary kinematic variable, and the associated total correlation matrix. One of the key advantages to the MAT is that, through the bookkeeping provided by the \mnvhnd, a user retains access to an observable in all systematic universes throughout every stage of their analysis. This enables, for example, detailed studies comparing the relative impact of uncertainties on different measurements.
  
%%%%%%%%%%%%%%%%%%%%%%%
%%%%%%%%%%%%%% Figure 2
%%%%%%%%%%%%%%%%%%%%%%%
%%%%%%%%%%%%%%%%%%%%%%%

\vspace{-1ex}
\section{Systematic Universe Classes}
\label{sec:universes_technical}
The MAT implements the many universe approach to error analysis, described in Section~\ref{sec:systematics}, using the \mnvhnd to store histograms and a \universe interface to abstract the concept of systematic universes. In a typical analysis, the user defines a CV \universe which includes methods to calculate all required event kinematics and weights (using information from the event record), and systematic \universe classes model alternatives to the default analysis by overriding the behavior of the  CV \universe. The methods of the CV \universe serve as a base interface for all other \universes, with each systematic \universe overriding one or more of the CV methods. In the case of vertical systematic shifts, one or more methods which calculate an event's weights are overridden, and in the case of lateral systematic shifts, one or more methods which calculate an event's kinematics are overridden. An analysis includes a loop over all \universes (CV and systematic), and the event kinematics and weights in each \universe are used to fill histograms in the corresponding \mnverrorbands of \mnvhnds. %They shift individual histograms within an \mnvhnd indirectly by changing how observables are calculated. MAT analysis programs are loops over \universe classes that separate core physics calculations from systematic uncertainty estimation.
%
%Universes as shifts
%The MAT abstracts each alternative calculation into a systematic \universe class. The CV analysis calculations become a base interface for all other \universes. A systematic universe overrides one or more CV calculation functions. Overriding part of an event's weight models a vertical systematic shift, and overriding an observable models a lateral shift.

% "tuple", "tree", "event record" <-- pick one.

Code Example \ref{lst:eventLoop} shows an example of an event loop in a MAT-based analysis. Within the loop over independent events is a loop over systematic \universes. In the MAT's simplest form, the full event selection is repeated for each \universe (see code starting on line 9). Compared to the CV analysis, each systematic \universe might, on an event-by-event basis, change the result of applying a cut (line 9) or calculating an event kinematic (line 10) or weight (line 11).

%%%%%%%%%%%%%%%%%%%%%%%
%%%%%%% Code Example 1
%%%%%%%%%%%%%%%%%%%%%%%
\noindent    % don't page break on code blocks 
\begin{minipage}{\linewidth}    % don't page break on code blocks
\begin{lstlisting}[style = mystyle, label=lst:eventLoop, caption={An example of an analysis event loop in the MAT. Within the loop over independent events is a loop over systematic \universes. In each \universe, cuts are applied and event kinematics and weights are calculated. The \passescuts function needs no special logic to handle systematic uncertainties that may affect it. Rather than fill a \thoned, the user instead fills an \mnvhoned.}]
double PassesCuts(const ExampleUniverse& univ) {
  return univ.GetLeptonEnergy() < 20;
}

MnvH1D energyHistogram(systematics);

for(const auto& event: eventRecord) {
  for(const ExampleUniverse& univ: systematics) {
    if(PassesCuts(univ)) {
      const double energy = univ.GetEnergy();
      const double weight = univ.GetWeight();
      energyHistogram.FillUniverse(univ, energy, weight);
    }
  }
}
\end{lstlisting}
\end{minipage}    % don't page break on code blocks 
%%%%%%%%%%%%%%%%%%%%%%%

Code Example \ref{lst:cvUniv} shows an example of a \universe. This class serves both as the interface that all systematics implement and as the CV implementation itself.  The MAT is written in C++, so weights and observables are virtual functions that can be overridden by derived classes. \lstinline[style=myStyle]{BaseUniverse} is the parent class to all \universes, and can provide experiment management and optimization tools.

%%%%%%%%%%%%%%%%%%%%%%%
%%%%%%% Code Example 2
%%%%%%%%%%%%%%%%%%%%%%%
\noindent    % don't page break on code blocks 
\begin{minipage}{\linewidth}    % don't page break on code blocks 
\begin{lstlisting}[style = mystyle, label=lst:cvUniv, caption={Example of a \universe, including methods to calculate event kinematics and weights. This class serves both as the \universe interface for an analysis and as the CV \universe implementation.  Every observable that could be affected by a systematic uncertainty is implemented as a virtual function.  An observable can be shifted by overriding its virtual function as done in Code Example \ref{lst:lateralUniv}.}] 
class ExampleUniverse: public BaseUniverse
{
  public:
    //Constructor and destructor details...
    
    virtual double GetLeptonEnergy() const {
      return m_tree->Get("lepton_energy");
    }
    
    virtual double GetRecoilEnergy() const {
      return m_tree->Get("recoil_energy");
    }
    
    virtual double GetEnergy() const {
      return GetLeptonEnergy() + GetRecoilEnergy(); 
    }
    
    virtual double GetTrackPIDScore() const {
        return m_tree->Get("track_PID_score");
    }
    
    virtual double GetWeight() const {
      return 1;
    }
    
  private:
    TreeWrapper* m_tree;
};
\end{lstlisting}
\end{minipage}    % don't page break on code blocks 
%%%%%%%%%%%%%%%%%%%%%%%

Systematic \universes are implemented as mixins, as in Code Example \ref{lst:verticalUniv}, so that uncertainties central to an experiment are implemented the same way by all analyzers. In this example, \lstinline[style=myStyle]{ResonantPionUniverse<>} is a template for a class that derives from a CV \universe with a \lstinline[style=myStyle]{GetWeight()} function. This class template only needs to know about the weight it shifts. When we construct the list of \universe pointers to loop over, the base \universe class carries the rest of the CV interface into \lstinline[style=myStyle]{ResonantPionUniverse<Universe>}. Similarly, in Code Example \ref{lst:lateralUniv} the \lstinline[style=myStyle]{GetLeptonEnergy()} observable is shifted while an arbitrary \universe interface is left intact. If these \universes were plugged into the \lstinline[style=myStyle]{systematics} container in Code Example \ref{lst:eventLoop}, \lstinline[style=myStyle]{CalibrationLeptonEnergyUniverse} (Code Example \ref{lst:lateralUniv}) would change which events make it past line 6, and \lstinline[style=myStyle]{ResonantPionUniverse} (Code Example \ref{lst:verticalUniv}) would change the shape of the histogram stored in the \thoned of the corresponding error band in \energyHistogram.  \energyHistogram itself is an \mnvhoned which contains the CV histogram and two \mnverrorbands.

%Why this design is a good idea
The separation of concerns between physics calculations and the evaluation of systematic uncertainties brings some broad advantages to MAT-based analyses.  An analysis routine that fills MnvHnDs is modeled by the same code for each \universe.  Analyzers only need to study one path of control to understand what quantity each histogram contains.  A developer focused on a particular systematic uncertainty only needs to read a \universe's implementation rather than hunt for blocks of code scattered throughout the analysis routine. Version control systems automatically provide a concise summary of how a systematic uncertainty algorithm has developed over time. Furthermore, any MAT analysis program is modular and does not need to be changed to introduce a new systematic uncertainty.

%%%%%%%%%%%%%%%%%%%%%%%
%%%%%%% Code Example 3
%%%%%%%%%%%%%%%%%%%%%%%
\noindent    % don't page break on code blocks 
\begin{minipage}{\linewidth}    % don't page break on code blocks 
\begin{lstlisting}[style = mystyle, label=lst:verticalUniv, caption=The event weight is scaled up 10\% for events which are truly resonant pion interactions in this \universe. This is an example of a "vertical" systematic shift.]
template <class UNIVERSE>
class ResonantPionUniverse: public UNIVERSE
{
  public:
    //Constructors to match UNIVERSE go here
  
    double GetWeight() const override {
      return this->IsResonantPion()?UNIVERSE::GetWeight():1.1*UNIVERSE::GetWeight();
    }
};
\end{lstlisting}
\end{minipage}    % don't page break on code blocks 
%%%%%%%%%%%%%%%%%%%%%%%

%%%%%%%%%%%%%%%%%%%%%%%
%%%%%%% Code Example 4
%%%%%%%%%%%%%%%%%%%%%%%
\noindent    % don't page break on code blocks 
\begin{minipage}{\linewidth}    % don't page break on code blocks 
\begin{lstlisting}[style = mystyle, label=lst:lateralUniv, caption=The event's lepton energy is increased by 50 MeV in this \universe. This is an example of a "lateral" systematic shift.]
template <class UNIVERSE>
class CalibrationLeptonEnergyUniverse: public UNIVERSE
{
  public:
    //Constructors to match UNIVERSE go here

    double GetLeptonEnergy() const override {
      return UNIVERSE::GetLeptonEnergy() + 50; //MeV
    }
};
\end{lstlisting}
\end{minipage}    % don't page break on code blocks 
%%%%%%%%%%%%%%%%%%%%%%%

\vspace{-1ex}
\section{MAT Adoption}
\label{sec:application}
The MAT's flagship tools, the \mnvhnd and \universe classes, are not reliant on particular technologies, but are merely powerful aids to many-universe binned analyses. They work well together, but are also standalone classes and can each be adopted individually to support an experiment's systematics infrastructure. The \mnvhnd on its own can simplify analysis calculations in many universes, provide on-demand covariance matrices, enable visualization of individual error bands (as shown in Figure \ref{fig:systematicsBreakdown}), and serve as a standard histogram container for experiments. The \universe classes can standardize systematics treatment across an experiment and isolate systematic impacts to the single simulation parameter or analysis quantity that they affect.

Beyond these tools, the MAT has an intended approach to analysis, which, when adopted, unlocks the full potential of the toolkit. This approach, demonstrated in Code Example~\ref{lst:eventLoop}, uses a single event-universe loop to measure an analysis quantity and collect into an \mnvhnd all information needed to calculate its systematic errors. This approach simplifies analysis flow, gives a central role to systematic uncertainties, and makes transparent the particular many-universe strategy employed. This approach, however, is best-used in concert with certain experiment-wide choices regarding data organization and analysis infrastructure design.

It would be difficult to capture the full range of possible conditions in an experiment's analysis environment and their consequences on MAT implementation, so we consider only \minerva here as an illustrative case. On \minerva, the common reconstruction pass over raw data is relatively minimal. Each analyzer thus has access to low-level variables, and runs their own reconstruction (for example, specialized tracking) on top of the general reconstruction to produce a custom tuple with new branches tailored to their analysis needs. This design choice is well-suited to the MAT approach because access to low-level objects allows users to dynamically study new sources of uncertainty that affect low- or high-level variables alike. In general, sources of systematic uncertainty can shift quantities throughout the simulation-reconstruction-analysis chain. Some are most conveniently expressed as a shift in a high-level, downstream analysis variable, which can be directly encoded in a \universe class (see, for example, Code Example~\ref{lst:lateralUniv}). Others may most conveniently shift a low-level simulation variable with consequences downstream on the analysis that are too-complicated to summarize in terms of high-level variables (see Code Example \ref{lst:lateralUniv2}). After quantifying the new source of uncertainty and the point in the simulation-reconstruction-analysis chain where it should be introduced, \minerva analyzers can then share across the experiment a corresponding \universe class and (if needed) a reconstruction prescription to produce systematically shifted higher-level branches.

%%%%%%%%%%%%%%%%%%%%%%%
%%%%%%% Code Example 5
%%%%%%%%%%%%%%%%%%%%%%%
\noindent    % don't page break on code blocks 
\begin{minipage}{\linewidth}    % don't page break on code blocks 
\begin{lstlisting}[style = mystyle, label=lst:lateralUniv2, caption={The uncertainty in target mass has a complex impact on a track's particle identification score. The shift is evaluated during reconstruction and its effect on particle identification is stored as a tuple branch, \lstinline{track_PID_score_target_mass_shift}.}]
template <class UNIVERSE>
class TargetMassUniverse: public UNIVERSE
{
  public:
    //Constructors to match UNIVERSE go here
    
    double GetTrackPIDScore() const override {
        return m_tree->Get("track_PID_score_target_mass_shift");
    }
};
\end{lstlisting}
\end{minipage}    % don't page break on code blocks 
%%%%%%%%%%%%%%%%%%%%%%%

\minerva's infrastructure choices, individual reconstruction and access to low-level objects, were feasible within \minerva's processing and data storage constraints, and they enable MAT's event-universe loop to be a practical one-stop-shop for filling the CV and systematically-altered histograms of the \mnvhnd. These choices further enable agile development of experiment-wide systematic errors in which users are able to express systematics in their most convenient form (as shifts in either high- or low-level variables).

It is not practical in every case to provide the user access to the lowest-level parameters which may be shifted by a source of systematic uncertainty. Interaction event generators, for example, provide event weights which are relatively high-level expressions of an underlying shift on a model parameter. \minerva has found a good compromise between providing users direct access to some low-level variables for studying new systematics and offering higher-level aggregated shifts as weights to avoid computationally expensive re-simulation. For any compromise, in order to make use of MAT's single-loop analysis approach, it is critical that experiments make available to the user a representation of a systematic shift that they can propagate through their analysis.

\vspace{-1ex}
\section{Conclusion}
\label{sec:conclusion}
% Summary of sections 1-5
In section \ref{sec:intro}, the need for common error analysis tools and a top-down approach to error analysis was motivated and the MAT was introduced. Section \ref{sec:systematics} introduced the abstract concept of universes and their role in the many universe error analysis method. Sections \ref{sec:mnvhnd} and \ref{sec:universes_technical} described MAT's foundational tools, the \mnvhnd and \universe classes, and it showed an example of typical MAT analysis flow. Finally, section \ref{sec:application} discussed the adoption of all or part of the MAT, and the important experiment-wide choices that impact its potential benefits.

There are extensions to the MAT which build on the functionality we have presented. In addition to providing the analysis foundation for dozens of \minerva cross section measurements, the MAT is now serving \minerva's data preservation effort. For eventual analysis of preserved \minerva data, measurements in terms of new observables is challenging because the expertise needed to assign systematic uncertainties to those observables may no longer exist. A MAT analysis solves this problem automatically for any combination of preserved observables. \minerva has also built on a central principle of the MAT and had some success recasting new models as reweights of existing models \cite{Rodrigues:2015hik}. MAT analyses can introduce reweights of the underlying model the same way a vertical systematic universe is implemented.

% Benefits of MAT in nu experiments
As experiments grow, accessible and flexible analysis systems will become increasingly important for maximizing physics output. The MAT lowers the barrier to entry for analyzers by providing them a working error analysis system, thereby allowing users to spend less time writing code and more time thinking about physics. MAT's systematics standardization further reduces code duplication and potential for mistakes.

% Future prospects for the MAT
The \minerva collaboration is preparing to publicly release the MAT this year, and continues active development to expand its capabilities. The MAT has not yet been used to perform a neutrino oscillation analysis or to apply a wide variety of machine learning methods, but we see no obstacles to those applications. Users of the MAT are currently exploring the idea of sharing systematics inter-experimentally. For example, we believe that broad adoption of the MAT could facilitate shared implementation of GENIE \cite{Andreopoulos:2009rq} model uncertainties in the neutrino interactions community. Outside of HEP, MAT's ideas could be implemented in new technologies and in any field which performs a many-universe binned analysis.
%% port to python

\bibliography{main}

\end{document}